\documentclass[11pt]{article}
\usepackage{bbm}
\usepackage{fullpage}
\usepackage[final]{graphicx}
\usepackage{amssymb}
\usepackage{amsmath}
\usepackage{amscd}
\usepackage{amsfonts}
\usepackage{amsthm}
\usepackage{color}
\usepackage{mathrsfs}

\usepackage{alltt, url}

\usepackage{graphicx}

\usepackage{algorithmic}
\usepackage[boxed, section]{algorithm}

\theoremstyle{plain}
\newtheorem{definition}{Definition}[section]

\theoremstyle{plain}
\newtheorem{theorem}[definition]{Theorem}

\theoremstyle{plain}
\newtheorem{corollary}[definition]{Corollary}

\theoremstyle{plain}
\newtheorem{lemma}[definition]{Lemma}

\theoremstyle{plain}

\begin {document}

 \title {An $O(\log k \log^2 n)$-competitive Randomized Algorithm for the $k$-Sever Problem}

\author {
    {Wenbin Chen}\footnote {Email:cwb2011@gzhu.edu.cn}
\thanks {Department of Computer Science, Guangzhou University, P.R. China}
\thanks {State Key Laboratory for Novel Software Technology, Nanjing
University, P.R. China}
}

\date {}

\maketitle
\begin {center}
\begin {minipage}{120mm}
\vskip 0.8in
\begin {center}{\bf Abstract} \end {center}
{\mbox {} \mbox {} \mbox {} \mbox {}\mbox{}


In this paper,   we show that there is an $O( \log k\log^2
n)$-competitive randomized  algorithm for the $k$-sever problem on any metric
space with $n$ points, which improved the
previous best competitive ratio $O( \log^2 k \log^3 n \log \log n)$
by Nikhil Bansal et al. (FOCS 2011, pages 267-276).

 }
\bigskip

\end {minipage}
\end {center}


\section{Introduction}


The $k$-sever problem is to schedule $k$ mobile servers to
serve a sequence of requests in a metric space with the minimum
possible movement distance. In 1990, Manasse et al. introduced the $k$-sever problem as a
generalization of several important online problems such as paging
and caching problems \cite {MMS} ( Its conference version is
\cite{MMS88}), in which they proposed a 2-competitive algorithm for
the 2-sever problem and a $n-1$-competitive algorithm for the $n-1$
sever problem in a $n$-point metric space. They still showed that
any deterministic online algorithm for the $k$-sever problem is of
competitive ratio at least $k$. They proposed the well-known
$k$-sever conjecture: for the $k$-sever problem on any metric space
with more than $k$ different points, there exists a deterministic
online algorithm with competitive ratio $k$.

It was in \cite{MMS} shown that the $k$-sever conjecture holds for
two special cases: $k=2$ and $n=k+1$. The $k$-sever conjecture also
holds for the $k$-sever problem on a uniform metric. The special
case of the $k$-sever problem on a uniform metric is called the
paging (also known as caching) problem. Slator and Tarjan have
proposed a $k$-competitive algorithm for the paging problem
\cite{ST}. For some other special metrics such as line, tree, there
existed $k$-competitive online algorithms. Yair Bartal and Elias
Koutsoupias show that the Work Function Algorithm for the $k$-sever
problem is of $k$-competitive ratio in the following special metric
spaces: the line, the star, and any metric space with $k+2$ points
\cite{BK}. Marek Chrobak and Lawrence L. Larmore proposed the
$k$-competitive Double-Coverage algorithm for the $k$-sever problem
on trees \cite{CL}.

For the $k$-sever problem on the general metric space, the $k$-sever
conjecture remain open. Fiat et al. were the first to show that
there exists an online algorithm of competitive ratio that depends
only on $k$ for any metric space: its competitive ratio is
$\Theta((k!)^3)$. The bound was improved later by Grove who showed
that the harmonic algorithm is of competitive ratio $O(k2^k)$
\cite{G}. The result was improved to $(2^k\log k)$ by Y.Bartal and
E. Grove \cite{BG}. A significant progress was achieved by
Koutsoupias and Papadimitriou, who proved that the work function
algorithm is of competitive ratio $2k-1$ \cite{KP}.

%

Generally, people believe that randomized online algorithms can produce better
competitive ratio than their deterministic counterparts. For
example, there are several $O(\log k)$-competitive algorithms for
the paging problem   and a $\Omega(\log k)$ lower bound on the
competitive ratio  in \cite{FKLM, MS, ACN, BBN07}. Although there
were much work \cite{BKRS, BBM, BLMN}, the $\Omega(\log k)$ lower
bound is still best lower bound in the
randomized case. 
Recently, N. Bansal et al. propose the first
polylogaritmic-competitive randomized algorithm for the $k$-sever
problem on a general metrics spcace \cite{BBMN}. Their randomized
algorithm is of competitive ratio $O( \log^2 k \log^3 n \log \log
n)$ for any metric space with $n$ points, which improves on the
deterministic $2k-1$ competitive ratio of Koutsoupias and
Papadimitriou whenever $n$ is sub-exponential $k$.

  For the $k$-server problem on the general
metric space, it is widely conjectured that there is an $O(\log
k)$-competitive randomized algorithm, which is called as the
randomized $k$-server conjecture. For the paging problem ( it
corresponds to the $k$-sever problem on a uniform metric), there is
 $O(\log k)$-competitive algorithms \cite{FKLM, MS, ACN}. For the
weighted paging problem ( it corresponds to the $k$-sever problem on
a weighted star metric space), there were also  $O(\log
k)$-competitive algorithms \cite{BBN07, BBN10} via the online
primal-dual method. More extensive literature on the $k$-server
problem can be found in \cite{K, BY}.

%
%

In this paper, we  show that there exists a
randomized $k$-sever algorithm of $O( \log k\log^2 n)$-competitive
ratio for any  metric space with $n$ points, which improved the
previous best competitive ratio $O( \log^2 k \log^3 n \log \log n)$
by Nikhil Bansal et al. \cite{BBMN}.

In order to get our results, we use the online primal-dual method,
which is developed by Buchbinder and Naor et al. in recent years.
Buchbinder and Naor et al. have used the primal-dual method to
design online algorithms for many online problems such as covering
and packing problems, the ad-auctions problem and so on \cite{BBN07,
BJN07, BN05, BN06,  BN09, BBN102}.
 First, we propose a primal-dual formulation for the fraction
$k$-sever problem on a weighted hierarchical well-separated tree (HST). Then, we design an $O(\ell\log k)$-competitive online
algorithm for the fraction $k$-sever problem on a weighted HST with depth $\ell$. Since any HST with $n$ leaves can be transformed
into a weighted HST with depth $O(\log n)$ with any leaf to leaf distance distorted by at most a constant \cite{BBMN}, thus, we get an $O(\log k\log n)$-competitive online
algorithm for the fraction $k$-sever problem on an HST. Based on the known
relationship between the fraction $k$-sever problem and the
randomized $k$-sever problem, we get that there is an $O( \log k\log
n)$-competitive randomized  algorithm for the $k$-sever problem on an HST with $n$ points.
By the metric embedding theory \cite{FRT}, we get that there is an $O( \log k\log^2
n)$-competitive randomized  algorithm for the $k$-sever problem on any metric space with $n$ points.


\section{Preliminaries}

In this section, we give some basic definitions.

\begin{definition}
(Competitive ratio adapted from \cite{T05}) For a deterministic
online algorithm DALG, we call it  $r$-competitive if there exists a
constant $c$ such that for any request sequence $\rho$,
$cost_{DALG}(\rho)\leq r\cdot cost_{OPT}(\rho) + c$, where
$cost_{DALG}(\rho)$ and $cost_{OPT}(\rho)$ are the costs of the
online algorithm $DALG$ and the best offline algorithm $OPT$
respectively.
\end{definition}

For a randomized online algorithm, we have a similar definition of
competitive ratio:

\begin{definition}
(Adapted from \cite{T05}) For a randomized online algorithm RALG, we
call it $r$-competitive if there exists a constant $c$ such that for
any request sequence $\rho$, $\mathbb{E}[cost_{RALG}(\rho)]\leq
r\cdot cost_{OPT}(\rho) + c$, where $\mathbb{E}[cost_{RALG}(\rho)]$
is the expected cost of the randomized online algorithm $RALG$.
\end{definition}

%
%
%
%


 In order
to analyze randomized algorithms for the $k$-sever problem, D.
T$\ddot{u}$rkou$\breve{g}$lu introduce the fractional $k$-sever
problem \cite{T05}.
On the fractional $k$-sever problem, severs are
viewed as fractional entities as opposed to units and an online
algorithm can move fractions of servers to the requested point.
%
%
%
%
%
%
%
%
%

\begin{definition}
(Fractional $k$-sever problem adapted from \cite{T05})  Suppose that
there are a metric space $\mathcal{S}$ and a total of $k$ fractional
severs located at the points of the metric space. Given a sequence of requests, each request must
be served by providing one unit server at requested point,  through moving fractional servers to the requested point. The
cost of an algorithm for servicing a sequence of requests is the
cumulative  sum of the distance incurred by each sever, where
moving a $w$ fraction of a server for a distance of $\delta$ costs
$w\delta$.

\end{definition}

In \cite{B96, B98}, Bartal introduce the definition of a {\it
Hierarchical Well-Separated Tree} (HST), into which a general metric
can be embedded  with a probability distribution.  For any internal node, the distance from
it to its parent node is $\sigma$ times of the distance from it to
its child node. The number $\sigma$ is called the stretch of the
HST. An HST with stretch $\sigma $ is called a $\sigma$-HST. In the following, we give
its formal definition.
%
\begin{definition}
(Hierarchically Well-Separated Trees (HSTs)\cite{CMP}). For $\sigma> 1$, a $\sigma$-Hierarchically Well-Separated
Tree ($\sigma$-HST) is a rooted tree $T = (V, E)$ whose edges length function $d$  satisfies the following properties:

(1). For any node $v$ and any two children $w_1, w_2$ of $v$, $d(v, w_1)=d(v, w_2)$.

(2). For any node $v$,  $d(p(v), v)=\sigma\cdot d(v, w)$, where $p(v)$ is the parent of $v$ and $w$ is a child of $v$.

(3). For any two leaves $v_1$ and $v_2$, $d(p(v_1), v_1)=d(p(v_2), v_2)$.

\end{definition}

Fakcharoenphol et al. showed the following result \cite{FRT}.

%

\begin{lemma}

If there is a $\gamma$-competitive randomized algorithm for the
$k$-sever problem on an $\sigma$-HST with all requests at the  $n$
leaves, then there exists an $O( \gamma \sigma \log n)$- competitive
randomized online algorithm for the $k$-server problem on any metric
space with $n$ points.


\end{lemma}

We still need the definition of a weighted hierarchically well-separated tree introduced in \cite{BBMN}.

\begin{definition}
(Weighted Hierarchically Well-Separated Trees (Weighted HSTs)\cite{BBMN}). A weighted $\sigma$-HST is a rooted tree satisfying the property (1), (3) of the definition 2.4 and  the property: $d(p(v), v)\geq\sigma\cdot d(v, w)$ for any node $v$ which is not any leaf  or the root, where $p(v)$ is the parent of $v$ and $w$ is a child of $v$.

\end{definition}

%
%
%
%

In \cite{BBMN}, Banasl et al. show that an arbitrary depth $\sigma$-HST with $n$ leaves can be embedded into an $O(\log n)$ depth weighted $\sigma$-HST with constant distortion, which is described as follows.

\begin{lemma}
Let $T$ be a $\sigma$-HST $T$ with $n$ leaves which is of possibly arbitrary depth. Then, $T$ can be transformed into a weighted $\sigma$-HST $\tilde{T}$ with depth $O(\log n)$ such that: the leaves of $\tilde{T}$ and $T$ are the same, and leaf to leaf distance in $T$ is distorted in $\tilde{T}$ by a at most $\frac{2\sigma}{\sigma-1}$ factor.
\end{lemma}

\section {An $O(\log^2 k)$ Randomized Algorithm for the $k$-Sever Problem on an HST when $n=k+1$}

In this paper, we view the $k$-sever problem as the weighed caching problem such that the cost  of evicting a page  out of the cache using another page satisfies the triangle inequality, i.e., a point is viewed as a page; the set of $k$ points that are served by $k$ severs is viewed as the cache which  holds $k$ pages; the distance of two points $i$ and $j$ is viewed the cost of evicting the corresponding  page $p_i$ out of the cache using the corresponding page $p_j$.


Let $[n]=\{p_1, \ldots, p_n\}$ denotes the set  of $n$ pages and $d(p_i, p_j)$ denotes the  cost of evicting the  page $p_i$ out of the cache using the page $p_j$ for any $p_i, p_j\in [n]$, which satisfies the triangle inequality: for any pages $i, j, s$, $d(p_i,p_i)=0; d(p_i,p_j)=d(p_j,p_i); d(p_i,p_j)\leq d(p_i,p_s)+d(p_s,p_j)$.
Let $p_1, p_2, \ldots,
p_M$ be the requested pages sequence until time $M$, where $p_t$ is
the requested page at time $t$.  At each time step, if the requested page $p_t$ is already in the cache then no cost is produced. Otherwise, the page $p_t$ must be fetched into the cache by evicting some other pages $p$ in the cache and  a cost $\sum d(p, p_t)$ is produced.

In this section, in order to clearly describe our algorithm design idea, we consider the case $n=k+1$.

First, we give some notations. Let $\sigma$-HST denote a hierarchically well-separated trees with stretch factor $\sigma$. Let $N$ be the number of nodes in a $\sigma$-HST and leaves be $p_1, p_2,\cdots, p_n$.  Let $\ell(v)$ denote the depth of a node $v$. Let $r$ denote the root node. Thus, $\ell(r)=0$. For any leaf $p$, let $\ell$ denote its depth, i.e. $\ell(p)=\ell$. Let $p(v)$ denote the parent node of a node $v$. $C(v)$ denote the set of children of a node $v$. Let $D$ denote the distance from the root to its a child. $D(v)$ denote the distance from a node $v$ to its parent, i.e. $D(v)=d(v, p(v))$. It is easy to know that $D(v)=\frac{D}{\sigma^{\ell (v)-1}}$. Let $T_v$ denote the subtree  rooted at $v$ and $L(T_v)$ denote the set of leaves in $T_v$. Let $|T_v|$ denote the number of the leaves in $T_v$. For a leaf $p_i$, let $A(p_i, j)$ denote the ancestor node of $p_i$ at the depth $j$. Thus, $A(p_i, \ell)$ is $p_i$,  $A(p_i, 0)$ is the root $r$ and so on. At time $t$, let variable $x_{p_i, t}$ denote the fraction of $p_i$ that is in the cache and $u_{p_i, t}$ denote the fraction of $p_i$ that is out of cache. Obviously, $x_{p_i, t}+u_{p_i, t}=1$ and $\sum\limits_{p\in [n]}x_{p,t}=k$. For a node $v$, let $u_{v,t}=\sum\limits_{p\in L(T_v)}u_{p,t}$, i.e., it is the total fraction of pages in the subtree $T_v$ which is out of the cache. It is easy to see that $u_{v,t}=\sum\limits_{w\in C(v)}u_{w,t}$. Suppose that at time 0, the
set of  initial $k$ pages in the cache is $I=\{p_{i_1},\ldots,
p_{i_k}\}$.

At time $t$, when the request $p_t$ arrives, if page $p_t$ is fetched mass $\Delta(p_t, p)$ into the cache by evicting out the page $p$ in the cache, then the evicting cost is $d(p, p_t)\cdot\Delta(p_t, p) $. For a $\sigma$-HST metric, suppose the path from $p_t$ to $p$ in it is: $p_t, v_j,\ldots, v_1, v, v'_1, \ldots, v'_j, p$, where $v$ is the first common ancestor node of $p_t$ and $p$. By the definition of a $\sigma$-HST, we have $D(p_t)=D(p)$ and $D(v_i)=D(v'_i)$ for any $1\leq i\leq j$.  Thus, $d(p, p_t)=D(p_t)+\sum\limits_{i=1}^{j}D(v_i)+D(p)+\sum\limits_{i=1}^{j}D(v'_i)=2D(p_t)+\sum\limits_{i=1}^{j}2D(v_i)$. So, the evicting cost is $(2D(p_t)+\sum\limits_{i=1}^{j}2D(v_i))\cdot \Delta(p_t, p)$. Since $p$ can be any page in $[n]\setminus\{p_t\}$, the evicting cost incurred at time $t$ is $\sum
\limits_{v=1}^{N}2D(v)\max \{0, u_{v,t-1}-u_{v,t}\}$.
 Thus,  we give the LP formulation for the fractional $k$-sever problem on a $\sigma$-HST as follows.

\smallskip

\begin{tabular}{|l|}\hline

(P) \hspace{1cm} Minimize $\sum \limits_{t=1}^{M}\sum
\limits_{v=1}^{N}2D(v)z_{v,t}+\sum\limits_{t=1}^{M}\infty\cdot
u_{p_t,t}$\\

Subject to $\forall t>0$ and $S\subseteq [n]$ with $|S|>k$,
$\sum\limits_{p\in
S}u_{p,t}\geq |S|-k$;\hspace{3.5cm }(3.1)\\

\mbox {} \mbox {} \hspace{1.5cm} $\forall t>0$ and a subtree $T_v (v\neq r)$, $z_{v,t}\geq \sum\limits_{p\in L(T_v)}(u_{p,t-1}-u_{p,t})$;\hspace{2cm }(3.2)\\

\mbox {} \mbox {} \hspace{1.5cm} $\forall t>0$ and node $v$,
$z_{v,t}, u_{v,t}\geq 0$;\hspace{6.7cm }(3.3)\\

\mbox {} \mbox {} \hspace{1.5cm} For $t=0$ and any leaf node $p\in I$,
$u_{p, 0}=0$;\hspace{4.8cm }(3.4)\\

\mbox {} \mbox {} \hspace{1.5cm} For $t=0$ and any leaf node $p\not\in
I$, $u_{p, 0}=1$;\hspace{4.8cm }(3.5)\\\hline

\end{tabular}

\smallskip

The first primal  constraint (3.1) states that at any time $t$, if
we take any set $S$ of vertices with $|S|>k$, then
$\sum\limits_{p\in S}u_{p,t}=|S|-\sum\limits_{p\in S}x_{p,t}\geq
|S|-\sum\limits_{p\in [n]}x_{p,t}= |S|-k$, i.e., the total number of pages out of the cache is at lease $|S|-k$. The variables $z_{v,t}$ denote the total fraction mass of pages in $T_v$ that are moved out of the subtree $T_v$ (Obviously, it is not needed to define a variable $z_{r,t}$ for the root node). The
fourth and fifth constraints ((3.4) and (3.5)) enforce the initial
 $k$ pages in the cache are $p_{i_1},\ldots, p_{i_k}$. The first
term in the object function is the sum of the moved cost out of the cache and the
second term enforces the requirement that the page $p_t$ must be in the cache at time $t$ (i.e., $u_{p_t,t}=0$).



\smallskip

Its dual formulation is as follows.

 \smallskip

\begin{tabular}{|l|}\hline

(D) \hspace{1cm} Maximize
$\sum\limits_{t=1}^{M}\sum\limits_{S\subseteq [n], |S|>k}
(|S|-k)a_{S,t}+\sum\limits_{p\not\in I} \gamma_p$\\

Subject to $\forall t$ and $p\in [n]\setminus \{ p_t$\}, $\sum\limits_{S:p\in S}
a_{S,t}-\sum\limits_{j=1}^{\ell}(b_{(A(p,j),t+1}-b_{(A(p, j),t})\leq 0$\hspace{1.1cm }(3.6)\\

\mbox {} \mbox {} \hspace{1.6cm}$\forall t=0$ and $\forall p\in [n]$,
$\gamma_p-\sum\limits_{j=1}^{\ell}b_{A(p, j),1}\leq 0$\hspace{5.3cm }(3.7)\\

\mbox {} \mbox {} \hspace{1.6cm}$\forall t>0$ and any subtree $T_v$, $b_{v,t}\leq 2D(v)$\hspace{5.5cm }(3.8)\\

\mbox {} \mbox {} \hspace{1.6cm}$\forall t>0$ and $v$ and $|S|>k$,
$a_{S,t}, b_{v,t}\geq 0$\hspace{5.7cm }(3.9)\\\hline

\end{tabular}

\smallskip

In the dual formulation, the variable $a_{S,t}$ corresponds to the
constraint of the type (3.1); the variable $b_{v,t}$ corresponds to
the constraint of the type (3.2); The variable $\gamma_p$
corresponds to the constraint of the type (3.4) and (3.5).


Based on above primal-dual formulation, we extend the design idea of Bansal et al.'s primal-dual algorithm for the metric task system problem on a $\sigma$-HST \cite{BBN102} to the $k$-sever problem on a $\sigma$-HST. The design idea of our online algorithm is described as follows. During the execution of our algorithm,  it always maintains the following  relation between the primal variable $u_{v, t}$ and dual variable $b_{v, t+1}$: $u_{v,t}=f(b_{v, t+1})=\frac{|T_v|}{k}(\exp(\frac{b_{v, t+1} }{2D(v)}\ln (1+k) )-1)$. When the request $p_t$ arrives at time $t$, the page $p_t$ is gradually fetched into the cache and other pages are gradually moved out of the cache by some rates until  $p_t$ is completely fetched into the cache(i.e, $u_{p_t, t}$ is decreased at some rate and other $u_{p,t}$ is increased at some rate for any $p\in [n]\setminus\{p_t\}$ until $u_{p_t, t}$ becomes 0). It can be viewed that we move mass $u_{p_t, t}$ out of leaf $p_t$ through its ancestor nodes and distribute it to other leaves  $p\in [n]\setminus \{p_t\}$. In order to compute the exact distributed amount at each  page $p\in [n]\setminus \{p_t\}$, the online algorithm should maintain the following invariants:






1. ( {\bf Satisfying Dual Constraints:}) It is tight for all dual constraints of type (3.6) on other leaves $[n]\setminus \{p_t\}$.


2. ({\bf  Node Identity Property:}) $u_{v, t}=\sum\limits_{w\in C(v)} u_{w, t}$ holds for each node $v$,.



  We give more clearer description of the online algorithm process. At time $t$, when the request $p_t$ arrives, we initially set $u_{p_t, t}=u_{p_t, t-1}$. If $u_{p_t, t}=0$, then we do nothing. Thus, the primal cost and the dual profit are both zero. All the invariants continue to hold.
  If $u_{p_t, t}\neq 0$, then we start to increase variable $a_S$ at rate $1$. At  each step, we would like to keep the dual
constraints (3.6) tight and maintain the node identity property. However, increasing variable $a_S$ violates the dual constraints(3.6)
on leaves in $[n]\setminus \{p_t\}$. Hence, we increase other dual variables  in order to keep these dual constraints (3.6)
tight. But, increasing these variables may also violate the node identity property. So, it makes us to
 update other dual variables. This process results in moving  initial  $u_{p_t, t}$ mass from
leaf $p_t$ to leaves $[n]\setminus \{p_t\}$. We stop the updating process when $u_{p_t, t}$ become $0$.

In the following, we will compute the exact rate at which we should move mass $u_{p_t, t}$ from $p_t$ through its ancestor nodes at time $t$ to other leaves in $[n]\setminus \{p_t\}$ in the $\sigma$-HST. Because of the space limit, we put  proofs of the following some claims in the Appendix. First, we show one property of the function $f$.

\begin{lemma}
$\frac{du_{v,t}}{db_{v, t+1}}=\frac{\ln (1+k)}{2D(v)}(u_{v,t}+\frac{|T_v|}{k})$.
\end{lemma}
\begin{proof}
Since $u_{v,t}=\frac{|T_v|}{k}(\exp(\frac{b_{v, t+1} }{2D(v)}\ln (1+k) )-1)$,  we take the derivative over $b_{v, t+1}$ and get the claim.
\end{proof}

In order to maintain the {\bf  Node Identity Property:} $u_{v, t}=\sum\limits_{w\in C(v)} u_{w, t}$  for each node $v$ at any time $t$, when $u_{v, t}$ is
increased or decreased, it is also required to increase or decrease the children of $v$ at some rate.  The connection between these rates is given.

\begin{lemma}
For a node $v$, if we increase variable $b_{v, t+1}$ at rate $h$, then we have the following equality:

$\frac{1}{\sigma}(u_{v,t}+\frac{|T_v|}{k})\cdot\frac{db_{v, t+1}}{dh}=\sum\limits_{w\in C(v)}\frac{db_{w, t+1}}{dh}\cdot(u_{w,t}+\frac{|T_v|}{k})$

\end{lemma}
%
%
%
%

We need one special case of lemma 3.2: when  the variable $b_{v, t+1}$ is increased (decreased) at rate $h$, it is required that the increasing (decreasing) rate of all children of $v$ is the same. By above lemma, we get:

\begin{lemma}
For  $v$ a node, assume that we increase (or decrease ) the variable $b_{v, t+1}$ at rate $h$. If the increasing (or decreasing ) rate of each $w\in C(v)$ is the same, then in order to keep the {\bf  Node Identity Property}, we should set the increasing (or decreasing ) rate for each child $w\in C(v)$ as follows:

$\frac{db_{w, t+1}}{dh}=\frac{1}{\sigma}\cdot\frac{db_{v, t+1}}{dh}$

\end{lemma}

%
%

Repeatedly applying this lemma, we get the following corollary.

\begin{corollary}
For a node $v$  with $\ell(v) = j$ and a path $P$ from leaf $p_i\in T_v$ to $v$, if  $b_{v,t+1}$ is increased ( or decreased)
at rate $h$ and the increasing (decreasing) rate of all children of any $v'\in P$ is the same, then $\sum\limits_{v'\in P}\frac{db_{v',t+1}}{dh}=\frac{db_{v,t+1}}{dh}\cdot \psi(j)$, where $\psi(j)=(1+\frac{1}{\sigma}+\frac{1}{\sigma^2}+\cdots+\frac{1}{\sigma^{\ell-j}})=(1+\Theta(\frac{1}{\sigma}))$.
\end{corollary}

We still require the following special case of lemma 3.2. Let $w_1$ be the first child of the node $v$. Assume that $b_{w_1, t+1}$ is
increased (or decreased) at some rate and the rate of increasing (or decreasing) $b_{w', t+1}$ is the same for every $w'\in C(v)$, $w'\neq w_1$.
If $b_{v, t+1}$ is unchanged, then the following claim should hold.

\begin{lemma}
Let $w_1, \ldots,w_m$ be the
children of a node $v$. Assume that we increase (or decrease) $w_1$ at rate $h$ and also increase  $w_2$ to $w_m$ at
the same rate $h$.  For $i\geq 2$, let $\frac{db_{w', t+1}}{dh}$ be $\frac{db_{w_i, t+1}}{dh}$. If we would like to maintain the amount $u_{v, t} $ unchanged,  then we should have:

$\frac{db_{w', t+1}}{dh}=\frac{u_{w_1,t}+\frac{|T(w_1)|}{k}}{u_{v,t}+\frac{|T(v)|}{k}}\cdot(-\frac{db_{w_1, t+1}}{dh}+\frac{db_{w', t+1}}{dh})$

\end{lemma}

%
%
%
%
%

\begin{theorem}
When request $p_t$ arrives at time $t$, in order to keep  the dual constraints tight and node identity property, if $a_{S,t}$ is increased  with rate $1$, we should decrease every $b_{A(p_t, j), t+1}$ ($1\leq j\leq \ell$) with rate:

 $\frac{ db_{A(p_t, j), t+1}}{da_{S,t}}=\frac{2+\frac{1}{n-1}}{\psi(j)}[(u_{A(p_t, j), t}+\frac{|T_{A(p_t, j)}|}{k})^{-1}-(u_{A(p_t, j-1), t}+\frac{|T_{A(p_t, j-1)}|}{k})^{-1}]$.


 For each sibling $w$ of $A(p_t, j)$, increase $b_{w, t+1}$ with the following rate:

 $\frac{db_{w, t+1}}{da_{S,t}}=\frac{2+\frac{1}{n-1}}{\psi(j)}(u_{A(p_t, j-1), t}+\frac{|T_{A(p_t, j-1)}|}{k})^{-1}$
\end{theorem}

 Thus,  we design an
online algorithm  for the fractional $k$-sever problem  as follows
(see Algorithm 3.1).

\begin{algorithm}
\caption{The online  primal-dual algorithm  for the fractional
$k$-sever problem on a $\sigma$-HST.} \label{alg}
\begin{algorithmic}[1]

\STATE At time $t=0$, we set $b_{p,1}=\gamma_p=0$ for all $p$ and set $b_{A(p, j), 1}=0$ for any $1\leq j\leq \ell$.

\STATE At time $t\geq 1$, when a request $p_t$ arrives:

\STATE Initially, we set $u_{p,t}=u_{p,t-1}$ for all $p$, and
$b_{p,t+1}$ is initialized to $b_{p,t}$.

%

%


\STATE  If $u_{p_t, t}=0$, then do nothing.

\STATE  Otherwise, do the following:

\STATE  Let $S=\{p: u_{p,t}<1\}$. Since $k=n-1$, $|S|>\sum\limits_{x\in S}x_{p,t}=k=n-1$. So, $S=[n]$.


\STATE While $u_{p_t,t}\neq0$:

\STATE \hspace{1.5cm} Increasing $a_{S,t}$ with rate $1$;

\STATE \hspace{1.5cm} For each $1\leq j\leq \ell$, decrease every $b_{A(p_t, j), t+1}$ with rate:

\STATE \hspace{1.5cm} $\frac{ db_{A(p_t, j), t+1}}{da_{S,t}}=\frac{2+\frac{1}{n-1}}{\psi(j)}[(u_{A(p_t, j), t}+\frac{|T_{A(p_t, j)}|}{k})^{-1}-(u_{A(p_t, j-1), t}+\frac{|T_{A(p_t, j-1)}|}{k})^{-1}]$,


\STATE \hspace{1.5cm} For each sibling $w$ of $A(p_t, j)$, increase $b_{w, t+1}$ with the following rate:

\STATE \hspace{1.5cm} $\frac{db_{w, t+1}}{da_{S,t}}=\frac{2+\frac{1}{n-1}}{\psi(j)}(u_{A(p_t, j-1), t}+\frac{|T_{A(p_t, j-1)}|}{k})^{-1}$

\STATE \hspace{1.5cm} For any node $v'$ in the path from $w$ to a leaf in $T_w$, if $w'$ be the child of $v'$, $\frac{db_{w', t+1}}{dh}=\frac{1}{\sigma}\cdot\frac{db_{v', t+1}}{dh}$

%
%
%
%

\end{algorithmic}
\end{algorithm}

\begin{theorem}
 \label{thm}
 The  online algorithm for the fractional $k$-sever problem on a $\sigma$-HST is of competitive ratio $15\ln^2 (1+k)$.
\end{theorem}

In \cite{T05}, Duru T$\ddot{u}$rko$\breve{g}$lu study the
relationship between fractional version  and randomized version of
the $k$-sever problem, which is given as follows.

\begin{lemma}

The fractional $k$-sever problem is equivalent to the randomized
$k$-sever problem on the line or circle, or if $k=2$ or $k=n-1$ for
arbitrary metric spaces.

\end{lemma}

Thus,  we get the following conclusion:

\begin{theorem}
There is a randomized algorithm with competitive ratio $15\ln^2(1+k)$
for the  $k$-sever problem on a $\sigma$-HST when
$n=k+1$.
\end{theorem}

By lemma 2.5, we get the following conclusion:

\begin{theorem}
 There is an $O(\log^2 k \log n)$ competitive randomized
algorithm for the $k$-sever problem on any metric space when $n=k+1$
.
\end{theorem}

\section {An $O(\ell\log k)$-competitive Fractional Algorithm for the $k$-Sever Problem on a Weighted HST with Depth $\ell$}

In this section, we first give an $O(\ell\log k)$-competitive fractional algorithm for the $k$-Sever problem on a weighted $\sigma$-HST with depth $\ell$.

We give another some  notations for a weighted HST. Let $\tilde{T}$ be a weighted $\sigma$-HST. For a node whose depth is $j$, let $\sigma_j=\frac{D(v)}{D(w)}=\frac{d(p(v),v)}{d(v, w)}$ where $w$ is a child of $v$. By the definition of a weighted $\sigma$-HST, $\sigma_j\geq \sigma$ for all $1\leq j \leq \ell-1$. For a node $v\in \tilde{T}$, if any leaf $p\in L(T(v))$ such that $u_{p, t}=1$, we call it a full node. By this definition, for a full node, $u_{v, t}=|L(T_v)|$. Otherwise, we call it non-full node. Let $NFC(v)$ is the set of non-full children node of $v$, i.e., $NFC(v)=\{w|w\in C(v)$ and $w$ is a non-full node $\}$. For a node $v$, let $NL(T_v)$ denote the set of non-full leaf nodes in $\tilde{T}_v$. Let $S=\{p|u_{p,t}<1\}$. Let  $P'$ denotes the path from $p_t$ to root $r$: $\{A(p_t, \ell)=P_t, A(p_t, \ell-1), \ldots, A(p_t, 1), \ldots, A(p_t, 0)=r\}$. For a node $v\in P'$, if there exists a $p\in S\setminus\{p_t\}$ such that $v$ is the first common ancestor of $p_t$ and $p$, we call it a common ancestor node in $P'$. Let $CA(p_t, S)$ denote the set of common ancestor nodes in $P'$. Suppose that $CA(p_t, S)=\{A(p_t, \ell_h), \ldots,  A(p_t, \ell_2), A(p_t, \ell_1)\}$, where $\ell_1< \ell_2<\ldots< \ell_h$. For a node $v$, let $u_{v, t}=\sum\limits_{p\in S} u_{p,t}$.  It is easy to know that $u_{v,t}=\sum\limits_{w\in NFC(v)}u_{w,t}$. Thus, for a full node $v$, $u_{v, t}=0$. For any $\ell_h< j< \ell, u_{A(p_t, j), t}=u_{A(p_t, \ell), t}$. For any $0\leq j< \ell_1, u_{A(p_t, 0), t}=u_{A(p_t, j), t}=u_{A(p_t, \ell_1), t}$. For any $\ell_{i-1}< j<\ell_i, u_{A(p_t, j), t}=u_{A(p_t, \ell_i), t}$.

The primal-dual formulation for the fractional $k$-sever problem on a weighted HST is the same as that on a HST in section 3. Based on the primal-dual formulation, the design idea of our online algorithm is similar to the design idea in section 3.   During the execution of our algorithm, it keeps the following  relation between the primal variable $u_{v, t}$ and dual variable $b_{v, t+1}$: $u_{v,t}=f(b_{v, t+1})=\frac{|NL(T_v)|}{k}(\exp(\frac{b_{v, t+1} }{2D(v)}\ln (1+k) )-1)$.
This relation  determines how much  mass of $u_{p_t, t}$ should be gradually moved out of leaf $p_t$  and
how it should be distributed among other leaves $S\setminus \{p_t\}$ until $p_t$ is completely fetched into the cache, i.e. $u_{p_t, t}=0$. Thus, at any time $t$, the algorithm maintains a distribution  $(u_{p_1, t}, \ldots, u_{p_n, t})$ on the leaves such that $\sum\limits_{p\in [n]}u_{p,t}=n-k$.

In order to compute the the exact rate at which we should move mass $u_{p_t, t}$ from $p_t$ through its ancestor nodes at time $t$ to other leaves $S\setminus \{p_t\}$ in the weighted $\sigma$-HST, using similar argument to that in section 3, we get following several claims. Because of the space limit, we put their proofs in the Appendix.

\begin{lemma}
$\frac{du_{v,t}}{db_{v, t+1}}=\frac{\ln (1+k)}{2D(v)}(u_{v,t}+\frac{|NL(T_v)|}{k})$.
\end{lemma}
\begin{proof}
Since $u_{v,t}=\frac{|NL(T_v)|}{k}(\exp(\frac{b_{v, t+1} }{2D(v)}\ln (1+k) )-1)$,  we take the derivative over $b_{v, t+1}$ and get the claim.
\end{proof}


\begin{lemma}
For a node $v$ with $\ell(v)=j$, if we increase variable $b_{v, t+1}$ at rate $h$, then we have the following equality:

$\frac{1}{\sigma_j}(u_{v,t}+\frac{|NL(T_v)|}{k})\cdot\frac{db_{v, t+1}}{dh}=\sum\limits_{w\in NFC(v)}\frac{db_{w, t+1}}{dh}\cdot(u_{w,t}+\frac{|NL(T_w)|}{k})$

\end{lemma}
%
%
%
%


\begin{lemma}
For  $v$ a node with $\ell(v)=j$, assume that we increase (or decrease ) the variable $b_{v, t+1}$ at rate $h$. If the increasing (or decreasing ) rate of each $w\in NFC(v)$ is the same, then in order to keep the {\bf  Node Identity Property}, we should set the increasing (or decreasing ) rate for each child $w\in NFC(v)$ as follows:

$\frac{db_{w, t+1}}{dh}=\frac{1}{\sigma_j}\cdot\frac{db_{v, t+1}}{dh}$

\end{lemma}

%
%

Repeatedly applying this lemma, we get the following corollary.

\begin{corollary}
For a node $v$  with $\ell(v) = j$ and a path $P$ from leaf $p_i\in T_v$ to $v$, if  $b_{v,t+1}$ is increased ( or decreased)
at rate $h$ and the increasing (decreasing) rate of all children of any $v'\in P$ is the same, then $\sum\limits_{v'\in P}\frac{db_{v',t+1}}{dh}=\frac{db_{v,t+1}}{dh}\cdot \phi(j)$, where $\phi(j)=(1+\frac{1}{\sigma_j}+\frac{1}{\sigma_j\cdot\sigma_{j+1}}+\cdots+\frac{1}{\sigma_j\cdot\sigma_{j+1}\ldots \sigma_{\ell -1}})\leq (1+\Theta(\frac{1}{\sigma}))$.
\end{corollary}


\begin{lemma}
Let $w_1, \ldots,w_m$ be the non-full
children node of a node $v$ (i.e., any $w_i\in NFC(v)$). Assume that we increase (or decrease) $w_1$ at rate $h$ and also increase  $w_2$ to $w_m$ at
the same rate $h$.  For $i\geq 2$, let $\frac{db_{w', t+1}}{dh}$ be $\frac{db_{w_i, t+1}}{dh}$. If we would like to maintain the amount $u_{v, t} $ unchanged,  then we should have:

$\frac{db_{w', t+1}}{dh}=\frac{u_{w_1,t}+\frac{|NL(T_{w_1})|}{k}}{u_{v,t}+\frac{|NL(T_v)|}{k}}\cdot(-\frac{db_{w_1, t+1}}{dh}+\frac{db_{w', t+1}}{dh})$

\end{lemma}

%
%
%



\begin{theorem}
When request $p_t$ arrives at time $t$, in order to keep  the dual constraints tight and node identity property, if $a_{S,t}$ is increased  with rate $1$, we should decrease every $b_{A(p_t, j), t+1}$ for each $j\in \{\ell_1+1, \ell_2+1, \ldots, \ell_h+1\}$ with rate:

 $\frac{ db_{A(p_t, j), t+1}}{da_{S,t}}=\frac{u_{r, t}+\frac{|S|}{k}}{\phi(j)}[(u_{A(p_t, j), t}+\frac{|NL(T_{A(p_t, j)})|}{k})^{-1}-(u_{A(p_t, j-1), t}+\frac{|NL(T_{A(p_t, j-1)})|}{k})^{-1}]$.


 For each sibling $w\in NFC(v)$ of $A(p_t, j)$, increase $b_{w, t+1}$ with the following rate:

 $\frac{db_{w, t+1}}{da_{S,t}}=\frac{u_{r, t}+\frac{|S|}{k}}{\phi(j)}(u_{A(p_t, j-1), t}+\frac{|NL(T_{A(p_t, j-1)})|}{k})^{-1}$
\end{theorem}

 Thus,  we design an
online algorithm  for the fractional $k$-sever problem on a weighted $\sigma$-HST as follows
(see Algorithm 4.1).

\begin{algorithm}
\caption{The online  primal-dual algorithm  for the fractional
$k$-sever problem on a weighted $\sigma$-HST.} \label{alg2}
\begin{algorithmic}[1]

\STATE At time $t=0$, we set $b_{p,1}=\gamma_p=0$ for all $p$.

\STATE At time $t\geq 1$, when a request $p_t$ arrives:

\STATE Initially, we set $u_{p,t}=u_{p,t-1}$ for all $p$, and
$b_{p,t+1}$ is initialized to $b_{p,t}$.

%

%


\STATE  If $u_{p_t, t}= 0$, then do nothing.

\STATE Otherwise, do the following:


\STATE  Let $S=\{p: u_{p,t}<1\}$. Suppose that $CA(p_t, S)=\{A(p_t, \ell_h), \ldots,  A(p_t, \ell_2), A(p_t, \ell_1)\}$, where $0\leq\ell_1< \ell_2<\ldots< \ell_h<\ell$.

\STATE While $u_{p_t, t}\neq 0$:

\STATE \hspace{1.5cm} Increasing $a_{S,t}$ with rate $1$;

\STATE \hspace{1.5cm} For each $j\in \{\ell_1+1, \ell_2+1, \ldots, \ell_h+1\}$, decrease every $b_{A(p_t, j), t+1}$ with rate:

\STATE \hspace{1.5cm} $\frac{ db_{A(p_t, j), t+1}}{da_{S,t}}=\frac{u_{r, t}+\frac{|S|}{k}}{\phi(j)}[(u_{A(p_t, j), t}+\frac{|NL(T_{A(p_t, j)})|}{k})^{-1}-(u_{A(p_t, j-1), t}+$

\STATE \hspace{1.5cm}$\frac{|NL(T_{A(p_t, j-1)})|}{k})^{-1}]$.

\STATE \hspace{1.5cm} For each sibling $w\in NFC(v)$ of $A(p_t, j)$, increase $b_{w, t+1}$ with the following rate:

\STATE \hspace{1.5cm} $\frac{db_{w, t+1}}{da_{S,t}}=\frac{u_{r, t}+\frac{|S|}{k}}{\phi(j)}(u_{A(p_t, j-1), t}+\frac{|NL(T_{A(p_t, j-1)})|}{k})^{-1}$

\STATE \hspace{1.5cm} For any node $v'$ in the path from $w$ to a leaf in $NL(T_w)$, if $w'\in NFC(v')$ and $\ell(v')=j$, $\frac{db_{w', t+1}}{dh}=\frac{1}{\sigma_j}\cdot\frac{db_{v', t+1}}{dh}$

%
%
%
%
\STATE \hspace{1.5cm}For $p\in S\setminus \{p_t\})$, if some $u_{p,t}$ reaches the value of $1$,
then we update $S\leftarrow S\setminus \{p\}$ and

\STATE \hspace{1.5cm}the set $NFC(v)$ for each ancestor node $v$ of $p$.

\end{algorithmic}
\end{algorithm}

\begin{theorem}
 \label{thm}
 The  online algorithm for the fractional $k$-sever problem on a weighted $\sigma$-HST with depth $\ell$ is of competitive ratio $4\ell\ln(1+k)$.
\end{theorem}

%
%
%
%
%


By lemma 2.7, we get:
\begin{theorem}
 There exists an $O(\log k\log n)$-competitive fractional algorithm for the $k$-sever problem on any $\sigma$-HST.
\end{theorem}

In \cite{BBMN}, Nikhil Bansal et al.
show the following conclusion.

\begin{lemma}
Let $T$ be a $\sigma$-HST with $\sigma>5$. Then any online
fractional $k$-sever algorithm on $T$ can be converted into a
randomized $k$-sever algorithm on $T$ with an $O(1)$ factor loss in
the competitive ratio.
\end{lemma}

Thus,  we get the following conclusion by Theorem 4.8:
\begin{theorem}
Let $T$ be a $\sigma$-HST with $\sigma>5$. There is a randomized
algorithm for the $k$-sever problem with a competitive ratio of
$O(\log k\log n)$ on $T$.

\end{theorem}

%
%

%

By lemma 2.5, we get the following conclusion:

\begin{theorem}
 For any metric space, there is a randomized
algorithm for the $k$-sever problem with a competitive ratio of
$O(\log k \log^2 n)$.
\end{theorem}

\section {Conclusion}
In this paper,  for any
 metric space with $n$ points, we show that there exist a randomized
 algorithm with $O( \log k \log^2 n)$-competitive ratio for the $k$-sever problem, which
improved the previous best competitive ratio $O( \log^2 k\log^3 n
\log \log n)$.

\section *{Acknowledgments}

\noindent We would like to thank the anonymous referees for their
careful readings of the manuscripts and many useful suggestions.

 Wenbin Chen's research has been partly supported  by the
National Natural Science Foundation of China (NSFC) under Grant
No.11271097., the research projects of Guangzhou education bureau
under Grant No. 2012A074. and  the project KFKT2012B01 from State
Key Laboratory for Novel Software Technology, Nanjing University.


\section*{Appendix}
%
%

{\bf Proofs for claims in section 3}

The proof for Lemma 3.2 is as follows:
\begin{proof}
 Since it is required to maintain $u_{v,t}=\sum\limits_{w\in C(v)}u_{w,t}$, we take the derivative of both sides and get that:

$\frac{du_{v,t}}{db_{v, t+1}}\cdot\frac{db_{v, t+1}}{dh} =\sum\limits_{w\in C(v)}\frac{du_{w,t}}{db_{w, t+1}}\cdot \frac{db_{w, t+1}}{dh}$.

By lemma 3.1, we get:
$\frac{\ln (1+k)}{2D(v)}(u_{v,t}+\frac{|T_v|}{k})=\sum\limits_{w\in C(v)}\frac{db_{w, t+1}}{dh}\cdot \frac{\ln (1+k)}{2D(w)}(u_{w,t}+\frac{|T_v|}{k})$.

Since $\frac{D(v)}{D(w)}=\sigma$, we get :

$\frac{1}{\sigma}(u_{v,t}+\frac{|T_v|}{k})\cdot\frac{db_{v, t+1}}{dh}=\sum\limits_{w\in C(v)}\frac{db_{w, t+1}}{dh}\cdot(u_{w,t}+\frac{|T_w|}{k})$
\end{proof}

The proof for Lemma 3.3 is as follows:
\begin{proof}
By above Lemma 3.2, if the increasing (or decreasing ) rate of each $w\in C(v)$ is the same, we get that:

 $\frac{1}{\sigma}(u_{v,t}+\frac{|T_v|}{k})\cdot\frac{db_{v, t+1}}{dh}=\frac{db_{w, t+1}}{dh}\cdot\sum\limits_{w\in C(v)}(u_{w,t}+\frac{|T_w|}{k})=\frac{db_{w, t+1}}{dh}\cdot(u_{v,t}+\frac{|T_v|}{k})$.

So, we get that: $\frac{db_{w, t+1}}{dh}=\frac{1}{\sigma}\cdot\frac{db_{v, t+1}}{dh}$
\end{proof}

The proof for Lemma 3.5 is as follows:
\begin{proof}
By lemma 3.2, in order to keep the amount $u_{v, t} $ unchanged, we get:

$\frac{db_{w_1, t+1}}{dh}\cdot (u_{w_1,t}+\frac{|T(w_1)|}{k})+\sum\limits_{w\in C(v)\setminus \{w_1\}}\frac{db_{w, t+1}}{dh}\cdot (u_{w,t}+\frac{|T(w)|}{k})=0$.

Thus, $(\frac{db_{w_1, t+1}}{dh}-\frac{db_{w', t+1}}{dh})\cdot (u_{w_1,t}+\frac{|T(w_1)|}{k})+\frac{db_{w', t+1}}{dh}\cdot\sum\limits_{w\in C(v)} (u_{w,t}+\frac{|T(w)|}{k})=0$

So, $(\frac{db_{w_1, t+1}}{dh}-\frac{db_{w', t+1}}{dh})\cdot (u_{w_1,t}+\frac{|T(w_1)|}{k})+\frac{db_{w', t+1}}{dh}\cdot (u_{v,t}+\frac{|T(v)|}{k})=0$

Hence, we get the claim.

\end{proof}

The proof for Theorem 3.6 is as follows:
\begin{proof}

When request $p_t$ arrives at time $t$,  we move mass $u_{p_t, t}$ from $p_t$ through its ancestor nodes to other leaves $[n]\setminus \{p_t\}$, i.e. $u_{p_t, t}$ is decreased and $u_{p,t}$ is increased for any $p\in [n]\setminus \{p_t\}$.  Since these mass moves out of each subtree $T_{A(p_t, j)}$ for each $1\leq j \leq \ell$, $u_{A(p_t, j), t}$ is decreased. By $u_{A(p_t, j), t}=f(b_{A(p_t, j), t+1})=\frac{|T_v|}{k}(\exp(\frac{b_{A(p_t, j), t+1} }{2D(v)}\ln (1+k) )-1)$ (we need to keep this relation during the algorithm), $b_{A(p_t, j), t+1}$ also decreases for each $1\leq j \leq \ell$. On the other hand, $u_{p,t}$ is increased for each $p\in [n]\setminus \{p_t\}$. Thus, for each node $v$ whose $T_v$ doesn't contain $p_t$, its mass $u_{v,t}$  is also increased. For each node $v$ whose $T_v$ doesn't contain $p_t$,  it must be a sibling of some node $A(p_t, j)$. For each $1\leq j \leq \ell$, we assume that all siblings $v'$ of  node $A(p_t, j)$ increase $b_{v', t+1}$ at the same rate.

In the following, we will compute  the increasing (or decreasing) rate of all dual variables in the $\sigma$-HST regarding $a_S$. For $1\leq j \leq \ell$, let $\nabla b_j=-\frac{b_{A(p_t, j), t+1}}{da_S}$ be the decreasing rate of $b_{A(p_t, j), t+1}$ regarding $a_S$. For $1\leq j \leq \ell$, let $\nabla b'_j=\frac{b_{w, t+1}}{da_S}$ be the increasing rate
 of $b_{w, t+1}$ for any siblings $w$ of $A(p_t, j)$  regarding $a_S$.

Using from top to down method, we can get a set of equations about the quantities $\nabla b_j$ and $\nabla b'_j$. First, we
consider the siblings of $A(p_t, 1)$ ( i.e. those nodes are children of root $r$, but they are not $A(p_t, 1)$). Let $w$ be one of
these siblings. If  $b_{w, t+1}$ is raised by $\nabla b'_1$,  by Corollary 3.4, the sum of $\nabla b'$ on any path from a leaf in $T_w$ to $w$ must be $\psi(1)\cdot\nabla b'_1$. Since $a_S$ is increasing with rate 1, it forces $\psi(1)\cdot\nabla b'_1=1$ in order to maintain the dual constraint (3) tight for leaves in $T_w$.

This considers the dual constraints for these leaves. Now, this increasing mass must
 be canceled out by decreasing the mass in $T_{A(p_t, 1)}$ since the mass $u_{r,t}$ in $T_r$ is not changed. Thus, in order to maintain the ``Node Identity Property" of root, by Lemma 3.5, we must set $\nabla b_1$ such that:

$\nabla b'_1=(\nabla b_1+\nabla b'_1)\cdot \frac{(u_{A(p_t, 1), t}+\frac{|T_{(A(p_t, 1))}|}{k})}{1+\frac{n}{n-1}}$

For siblings of node $A(p_t, 2)$, we use the similar  argument. Let $w$ be a sibling of $A(p_t, 2)$. Consider a path from
a leaf in $T_w$ to the $w$. Their dual constraint (3) already grows at rate $1+\psi(1)\nabla b_1$. This
must be canceled out by increasing $b_{w, t+1}$, and  if  $b_{w, t+1}$ is raised by $\nabla b'_2$,  by corollary 3.4, the sum of $\nabla b'$ on any path from a leaf in $T_w$ to $w$ must be $\psi(2)\cdot\nabla b'_2$. Thus, $\nabla b'_2$ must be set such that: $\psi(2)\cdot \nabla b'_2=1+\psi(1)\cdot \nabla b_1$

Again, this increasing mass must be canceled out by decreasing the mass
in $T_{A(p_t, 2)}$ . In order to keep the ``Node Identity Property" of $A(p_t, 1)$, by Lemma 3.5. we must set $\nabla b_2$  such that:
$\nabla b'_2=(\nabla b_2+\nabla b'_2)\cdot \frac{(u_{A(p_t, 2), t}+\frac{|T_{A(p_t, 2)}|}{k})}{(u_{A(p_t, 1), t}+\frac{|T_{(A(p_t, 1)}|}{k})}$

Continuing this method, we obtain a system of linear equations about all $\nabla b_j$ and  $\nabla b'_j$ ($1\leq j\leq \ell$). For maintaining the dual constraints tight, we get the following equations:

$\psi(1)\cdot \nabla b'_1=1$

$\psi(2)\cdot \nabla b'_2=1+\psi(1)\cdot \nabla b_1$

$\vdots$

$\psi(\ell)\cdot \nabla b'_{\ell}=1+\sum\limits_{i=1}^{\ell-1}\psi(i) \nabla b_i$

For keeping the node identity property, we get the following equations:

$\nabla b'_1=(\nabla b_1+\nabla b'_1)\cdot \frac{(u_{A(p_t, 1), t}+\frac{|T_{(A(p_t, 1))}|}{k})}{1+\frac{n}{n-1}}$

$\nabla b'_2=(\nabla b_2+\nabla b'_2)\cdot \frac{(u_{A(p_t, 2), t}+\frac{|T_{(A(p_t, 2))}|}{k})}{(u_{A(p_t, 1), t}+\frac{|T_{(A(p_t, 1))}|}{k})}$

$\vdots$

$\nabla b'_{\ell}=(\nabla b_{\ell}+\nabla b'_{\ell})\cdot \frac{(u_{A(p_t, {\ell}), t}+\frac{|T_{(A(p_t, {\ell}))}|}{k})}{(u_{A(p_t, \ell-1), t}+\frac{|T_{(A(p_t, \ell-1))}|}{k})}$.

We continue to solve the system of linear equations.

For each $1\leq j\leq \ell$,

$\psi(j)\cdot \nabla b'_{j}=1+\sum\limits_{i=1}^{j-1}\psi(i) \nabla b_i$

$=1+\sum\limits_{i=1}^{j-2}\psi(i) \nabla b_i+\psi(j-1) \nabla b_{j-1}$

$=\psi(j-1)\cdot \nabla b'_{j-1}+\psi(j-1) \nabla b_{j-1}$

$=\psi(j-1)\cdot (\nabla b'_{j-1}+ \nabla b_{j-1})$.

%
%
%
%


Since $\nabla b'_{j-1}=(\nabla b_{j-1}+\nabla b'_{j-1})\cdot \frac{(u_{A(p_t, {j-1}), t}+\frac{|T_{(A(p_t, {j-1}))}|}{k})}{(u_{A(p_t, j-2), t}+\frac{|T_{(A(p_t, j-2))}|}{k})}$, we get:

$\psi(j)\nabla b'_j=\psi(j-1)\nabla b'_{j-1}\cdot \frac{(u_{A(p_t, j-2), t}+\frac{|T_{(A(p_t, j-2))}|}{k})}{(u_{A(p_t, j-1), t}+\frac{|T_{(A(p_t, j-1))}|}{k})}$

Solving the recursion, we get:

$\nabla b'_j=\frac{1+\frac{n}{n-1}}{\psi(j)\cdot (u_{A(p_t, j-1), t}+\frac{|T_{(A(p_t, j-1))}|}{k})}$

$\nabla b_j=\frac{1+\frac{n}{n-1}}{\psi(j)}\cdot [(u_{A(p_t, j), t}+\frac{|T_{(A(p_t, j))}|}{k})^{-1}-(u_{A(p_t, j-1), t}+\frac{|T_{(A(p_t, j-1))}|}{k})^{-1}]$

\end{proof}

The proof for Theorem 3.7 is as follows:
\begin{proof}

Let $P$ denote the value of the objective function of the primal
solution and $D$ denote the value of the objective function of the
dual solution. Initially, let $P=0$ and $D=0$. In the following, we
prove three claims:

(1) The primal solution produced by the algorithm is feasible.

(2)  The dual  solution produced by the algorithm is feasible.

(3) $ P\leq 15\ln^2 (1+k)D$.

By three claims and weak duality of linear programs, the theorem
follows immediately.

First, we prove the claim (1) as follows. At any time $t$, since $S=[n]$ and the algorithm keeps $\sum\limits_{p\in [n]}=n-k=1=|S|-k$. So, the primal constraints (3.1) are satisfied.

%
%
%
%
%
%
%
%
%
%
%
%

Second, we prove the claim (2) as follows. By theorem 3.6, the dual constraints (3.6) are satisfied. Obviously, dual constraints (3.7) are satisfied.
For any node $v$, if $b_{v,t+1}=0$, then $u_{v,t}=0$; if $b_{v,t+1}=2D(v)$, then $u_{v,t}=|T_v|$. Thus, the dual constraints (3.8) are satisfied.



Third, we prove claim (3) as follows.  If  the algorithm increases
the variables $a_{S,t}$ at some time $t$, then: $\frac{\partial
D}{\partial a_{S,t}}=|S|-k=n-(n-1)=1$. Let's compute the primal cost. At depth $j (1\leq j\leq \ell)$, we compute the movement cost of our algorithm by the change of $\nabla b'_j$ as follows.


%


$\sum\limits_{w\in C(A(p_t, j-1))\backslash
\{A(p_t, j)\}}2D(w)\cdot \frac{du_{w,t}}{db_j'}\cdot \frac{db_j'}{a_S}$

$\hspace{1.0cm}=\nabla b'_j\cdot\sum\limits_{w\in C(A(p_t, j-1))\backslash
\{A(p_t, j)\}}2D(w)\cdot\frac{\ln(1+k)}{2D(w)}(u_{w,t}+\frac{|T_w|}{k})$

$\hspace{1.0cm}=\frac{2+\frac{1}{n-1}}{\psi(j)}\cdot\ln(1+k)\cdot\frac{\sum\limits_{w\in C(A(p_t, j-1))\backslash
\{A(p_t, j)\}}(u_{w,t}+\frac{|T_w|}{k})}{u_{A(p_t, j-1),t}+\frac{|T_{A(p_t, j-1)}|}{k}}$

$\hspace{1.0cm}\leq \frac{5}{2}\cdot\ln(1+k)\cdot \frac{\sum\limits_{w\in C(A(p_t, j-1))\backslash
\{A(p_t, j)\}}(u_{w,t}+\frac{|T_w|}{k})}{u_{A(p_t, j-1),t}+\frac{|T_{A(p_t, j-1)}|}{k}}$


Let $B_j$ denote $u_{A(p_t, j),t}+\frac{|T_{A(p_t, j)}|}{k}$. Then $\sum\limits_{w\in C(A(p_t, j-1))\backslash
\{A(p_t, j)\}}(u_{w,t}+\frac{|T_w|}{k})=B_{j-1}-B_j$.

Hence, the total movement cost over all $\ell$ levels is

$\frac{5}{2}\ln(1+k)\cdot \sum_{j=1}^{\ell}\frac{B_{j-1}-B_j}{B_{j-1}}$

$=\frac{5}{2}\ln(1+k)\cdot \sum_{j=1}^{\ell}(1-\frac{B_j}{B_{j-1}})$

$\leq \frac{5}{2}\ln(1+k)\cdot \sum_{j=1}^{\ell} \ln \frac{B_j}{B_{j-1}}$

$=\frac{5}{2}\ln(1+k)\cdot\ln \frac{B_0}{B_{\ell}}$

$= \frac{5}{2}\ln(1+k)\cdot\ln \frac{B_0}{u_{p_t, t}+\frac{1}{k}}$

$\leq \frac{5}{2}\ln(1+k)\cdot\ln kB_0$

$\leq \frac{5}{2}\ln(1+k)\cdot2\ln k\cdot B_0$

$\leq \frac{5}{2}\ln(1+k)\cdot2\ln k\cdot (2+\frac{1}{n-1})$

$= 15\ln^2(1+k)$

%
%
%
%
%
%
%

where the first inequality holds since $1\leq y-\ln y$ for any $0\leq y\leq 1$.

Thus, we get $ P\leq 15\ln^2(1+k)D$.

Let $OPT$ be the cost of the best offline algorithm. $P_{min}$ be
the optimal primal solution and $D_{max}$ be the optimal dual
solution. Then, $P_{min}\leq OPT$ since $OPT$ is a feasible solution
for the primal program. Based on the weak duality, $D_{max}\leq
P_{min}$. Hence, $\frac{P}{OPT}\leq \frac{P}{P_{min}}\leq
\frac{15\ln^2(1+k)D}{P_{min}}\leq
\frac{15\ln^2(1+k)D_{max}}{P_{min}}\leq\frac{15\ln^2(1+k)P_{min}}{P_{min}}
=15\ln^2(1+k)$.

So, the competitive ratio of this algorithm is  $15\ln^2(1+k)$.

%
%
%
%
%

\end{proof}

{\bf Proofs for claims in section 4}

The proof for Lemma 4.2 is as follows:
\begin{proof}
 Since it is required to maintain $u_{v,t}=\sum\limits_{w\in NFC(v)}u_{w,t}$, we take the derivative of both sides and get that:

$\frac{du_{v,t}}{db_{v, t+1}}\cdot\frac{db_{v, t+1}}{dh} =\sum\limits_{w\in NFC(v)}\frac{du_{w,t}}{db_{w, t+1}}\cdot \frac{db_{w, t+1}}{dh}$.

By lemma 3.1, we get:
$\frac{\ln (1+k)}{2D(v)}(u_{v,t}+\frac{|NL(T_v)|}{k})=\sum\limits_{w\in NFC(v)}\frac{db_{w, t+1}}{dh}\cdot \frac{\ln (1+k)}{2D(w)}(u_{w,t}+\frac{|NL(T_w)|}{k})$.

Since $\frac{D(v)}{D(w)}=\sigma_j$, we get :

$\frac{1}{\sigma_j}(u_{v,t}+\frac{|NL(T_v)|}{k})\cdot\frac{db_{v, t+1}}{dh}=\sum\limits_{w\in NFC(v)}\frac{db_{w, t+1}}{dh}\cdot(u_{w,t}+\frac{|NL(T_w)|}{k})$
\end{proof}

%
%
%

The proof for Lemma 4.3 is as follows:

\begin{proof}
By  Lemma 4.2, if the increasing (or decreasing ) rate of each $w\in NFC(v)$ is the same, we get that:

 $\frac{1}{\sigma_j}(u_{v,t}+\frac{|NL(T_v)|}{k})\cdot\frac{db_{v, t+1}}{dh}=\frac{db_{w, t+1}}{dh}\cdot\sum\limits_{w\in NFC(v)}(u_{w,t}+\frac{|NL(T_w)|}{k})=\frac{db_{w, t+1}}{dh}\cdot(u_{v,t}+\frac{|NL(T_v)|}{k})$.

So, we get that: $\frac{db_{w, t+1}}{dh}=\frac{1}{\sigma_j}\cdot\frac{db_{v, t+1}}{dh}$
\end{proof}

%
%
%

The proof for Lemma 4.5 is as follows:

%
%

\begin{proof}
By lemma 4.2, in order to keep the amount $u_{v, t} $ unchanged, we get:

$\frac{db_{w_1, t+1}}{dh}\cdot (u_{w_1,t}+\frac{|NL(T_{w_1})|}{k})+\sum\limits_{w\in NFC(v)\setminus \{w_1\}}\frac{db_{w, t+1}}{dh}\cdot (u_{w,t}+\frac{|NL(T_w)|}{k})=0$.

Thus, $(\frac{db_{w_1, t+1}}{dh}-\frac{db_{w', t+1}}{dh})\cdot (u_{w_1,t}+\frac{|NL(T_{w_1})|}{k})+\frac{db_{w', t+1}}{dh}\cdot\sum\limits_{w\in NFC(v)} (u_{w,t}+\frac{|NL(T_w)|}{k})=0$

So, $(\frac{db_{w_1, t+1}}{dh}-\frac{db_{w', t+1}}{dh})\cdot (u_{w_1,t}+\frac{|NL(T_{w_1})|}{k})+\frac{db_{w', t+1}}{dh}\cdot (u_{v,t}+\frac{|NL(T_v)|}{k})=0$

Hence, we get the claim.

\end{proof}

%
%
%
%

The proof for Theorem 4.6 is as follows:

\begin{proof}

When request $p_t$ arrives at time $t$,  we move mass $u_{p_t, t}$ from $p_t$ through its ancestor nodes to other non-full leaves nodes $S\setminus \{p_t\}$, i.e. $u_{p_t, t}$ is decreased and $u_{p,t}$ is increased for any $p\in S\setminus \{p_t\}$.  Since these mass moves out of each subtree $T_{A(p_t, j)}$ for each $1\leq j \leq \ell$, $u_{A(p_t, j), t}$ is decreased. By $u_{A(p_t, j), t}=f(b_{A(p_t, j), t+1})=\frac{|NL(T_v)|}{k}(\exp(\frac{b_{A(p_t, j), t+1} }{2D(v)}\ln (1+k) )-1)$ (we need to keep this relation during the algorithm), $b_{A(p_t, j), t+1}$ also decreases for each $1\leq j \leq \ell$. On the other hand, $u_{p,t}$ is increased for each $p\in S\setminus \{p_t\}$. Thus, for each non-full node $v$ whose $T_v$ doesn't contain $p_t$, its mass $u_{v,t}$  is also increased. For each non-full node $v$ whose $T_v$ doesn't contain $p_t$,  it must be a sibling of some node $A(p_t, j)$ where $j\in \{\ell_1, \ldots, \ell_h\}$. We assume that all siblings $v'$ of any node $v$ increase $b_{v', t+1}$ at the same rate

In the following, we will compute  the increasing (or decreasing) rate of all dual variables in the weighted $\sigma$-HST regarding $a_S$. For $1\leq j \leq \ell$, let $\nabla b_j=-\frac{b_{A(p_t, j), t+1}}{da_S}$ be the decreasing rate of $b_{A(p_t, j), t+1}$ regarding $a_S$. For each $j\in \{\ell_1+1, \ell_2+1, \ldots, \ell_h+1\}$, let $\nabla b'_j=\frac{b_{w, t+1}}{da_S}$ be the increasing rate
 of $b_{w, t+1}$ for any siblings $w\in NFC(A(p_t, j))$ of $A(p_t, j)$  regarding $a_S$.

Using from top to down method, we can get a set of equations about the quantities $\nabla b_j$ and $\nabla b'_j$.
 First, we
consider the siblings of $A(p_t,\ell_1+1)$ ( i.e. those nodes are children of $A(p_t,\ell_1)$, but they are not $A(p_t, \ell_1+1)$). Let $w$ be one of
these siblings. If  $b_{w, t+1}$ is raised by $\nabla b'_{\ell_1+1}$,  by Corollary 4.4, the sum of $\nabla b'$ on any path from a leaf in $T_w$ to $w$ must be $\phi(\ell_1+1)\cdot\nabla b'_{\ell_1+1}$. Since $a_S$ is increasing with rate 1, it forces $\phi(\ell_1+1)\cdot\nabla b'_{\ell_1+1}=1$ in order to maintain the dual constraint (3) tight for non-full leaf nodes in $T_w$.

This considers the dual constraints for these non-full leaf nodes. Now, this increasing mass must
 be canceled out by decreasing the mass in $T_{A(p_t, \ell_1+1)}$ since the mass $u_{A(p_t,\ell_1),t}$ in $T_{A(p_t,\ell_1)}$ is not changed. Thus, in order to maintain the ``Node Identity Property" of $A(p_t,\ell_1)$, by Lemma 4.5, we must set $\nabla b_1$ such that:

$\nabla b'_{\ell_1+1}=(\nabla b_{\ell_1+1}+\nabla b'_{\ell_1+1})\cdot \frac{(u_{A(p_t, \ell_1+1), t}+\frac{|NL(T_{(A(p_t, \ell_1+1))})|}{k})}{u_{A(p_t,\ell_1),t}+\frac{|S|}{k}}$

$=(\nabla b_{\ell_1+1}+\nabla b'_{\ell_1+1})\cdot \frac{(u_{A(p_t, \ell_2), t}+\frac{|NL(T_{(A(p_t, \ell_2))})|}{k})}{u_{r}+\frac{|S|}{k}}$

For siblings of node $A(p_t, \ell_2+1)$, we use the similar  argument. Let $w$ be a sibling of $A(p_t, \ell_2+1)$. Consider a path from
a non-full leaf node in $T_w$ to the $w$. Their dual constraint (3.6) already grows at rate $1+\phi(\ell_1+1)\nabla b_{\ell_1+1}$. This
must be canceled out by increasing $b_{w, t+1}$, and  if  $b_{w, t+1}$ is raised by $\nabla b'_{\ell_2+1}$,  by corollary 4.4, the sum of $\nabla b'$ on any path from a leaf in $T_w$ to $w$ must be $\phi(2)\cdot\nabla b'_2$. Thus, $\nabla b'_2$ must be set such that: $\phi(\ell_2+1)\cdot \nabla b'_{\ell_2+1}=1+\phi(\ell_1+1)\cdot \nabla b_{\ell_1+1}$.

Again, this increasing mass must be canceled out by decreasing the mass
in $T_{A(p_t, \ell_2)}$ . In order to keep the ``Node Identity Property" of $A(p_t, \ell_2)$, by Lemma 4.5. we must set $\nabla b_{\ell_2+1}$  such that:
$\nabla b'_{\ell_2+1}=(\nabla b_{\ell_2+1}+\nabla b'_{\ell_2+1})\cdot \frac{(u_{A(p_t, {\ell_2+1}), t}+\frac{|NL(T_{A(p_t, {\ell_2+1})})|}{k})}{(u_{A(p_t,\ell_2), t}+\frac{|NL(T_{(A(p_t,\ell_2)})|}{k})}$

$=(\nabla b_{\ell_2+1}+\nabla b'_{\ell_2+1})\cdot \frac{(u_{A(p_t, {\ell_3}), t}+\frac{|NL(T_{A(p_t, {\ell_3})})|}{k})}{(u_{A(p_t,\ell_2), t}+\frac{|NL(T_{(A(p_t,\ell_2)})|}{k})}$

Continuing this method, we obtain a system of linear equations about all $\nabla b_j$ and  $\nabla b'_j$ ($j\in \{\ell_1+1, \ell_2+1, \ldots, \ell_h+1\}$).
For maintaining the dual constraints tight, we get the following equations:

$\phi(\ell_1+1)\cdot \nabla b'_{\ell_1+1}=1$

$\phi(\ell_2+1)\cdot \nabla b'_{\ell_2+1}=1+\phi(\ell_1+1)\cdot \nabla b_{\ell_1+1}$

$\vdots$

$\psi(\ell_h+1)\cdot \nabla b'_{\ell_h+1}=1+\sum\limits_{i=1}^{h-1}\phi(\ell_i+1) \nabla b_{\ell_i+1}$

For keeping the node identity property, we get the following equations:

$\nabla b'_{\ell_1+1}=(\nabla b_{\ell_1+1}+\nabla b'_{\ell_1+1})\cdot \frac{(u_{A(p_t, \ell_2), t}+\frac{|NL(T_{(A(p_t, 1))})|}{k})}{u_r+\frac{|S|}{k}}$

$\nabla b'_{\ell_2+1}=(\nabla b_{\ell_2+1}+\nabla b'_{\ell_2+1})\cdot \frac{(u_{A(p_t, \ell_3), t}+\frac{|NL(T_{(A(p_t, \ell_3))})|}{k})}{(u_{A(p_t, \ell_2), t}+\frac{|NL(T_{(A(p_t, \ell_2))})|}{k})}$

$\vdots$

$\nabla b'_{\ell_h+1}=(\nabla b_{\ell_h+1}+\nabla b'_{\ell_h+1})\cdot \frac{(u_{A(p_t, {\ell}), t}+\frac{|NL(T_{(A(p_t, {\ell}))})|}{k})}{(u_{A(p_t, \ell_h), t}+\frac{|NL(T_{(A(p_t, \ell_h))})|}{k})}$.

We continue to solve the system of linear equations.

For each $1\leq j\leq h$,

$\phi(\ell_j+1)\cdot \nabla b'_{\ell_j+1}=1+\sum\limits_{i=1}^{j-1}\phi(\ell_i+1) \nabla b_{\ell_i+1}$

$=1+\sum\limits_{i=1}^{j-2}\phi(\ell_i+1) \nabla b_{\ell_i+1}+\phi(\ell_{j-1}+1) \nabla b_{\ell_{j-1}+1}$

$=\phi(\ell_{j-1}+1)\cdot \nabla b'_{\ell_{j-1}+1}+\phi(j-1) \nabla b_{\ell_{j-1}+1}$

$=\phi(\ell_{j-1}+1)\cdot (\nabla b'_{\ell_{j-1}+1}+ \nabla b_{\ell_{j-1}+1})$.

%
%
%
%


Since $\nabla b'_{\ell_{j-1}+1}=(\nabla b_{\ell_{j-1}+1}+\nabla b'_{\ell_{j-1}+1})\cdot \frac{(u_{A(p_t, \ell_j), t}+\frac{|NL(T_{(A(p_t, \ell_j))})|}{k})}{(u_{A(p_t, \ell_{j-1}), t}+\frac{|NL(T_{(A(p_t, \ell_{j-1}))})|}{k})}$, we get:

$\psi(\ell_j+1)\nabla b'_{\ell_j+1}=\psi(\ell_{j-1}+1)\nabla b'_{\ell_{j-1}+1}\cdot \frac{(u_{A(p_t, \ell_{j-1}), t}+\frac{|NL(T_{(A(p_t, \ell_{j-1}))})|}{k})}{(u_{A(p_t, \ell_j), t}+\frac{|NL(T_{(A(p_t, \ell_j))})|}{k})}$

Solving the recursion, we get:

$\nabla b'_{\ell_j+1}=\frac{u_r+\frac{|S|}{k}}{\phi(\ell_j+1)\cdot (u_{A(p_t, \ell_j), t}+\frac{|NL(T_{(A(p_t, \ell_j))})|}{k})}$

$\nabla b_{\ell_j+1}=\frac{u_r+\frac{|S|}{k}}{\phi({\ell_j+1})}\cdot [(u_{A(p_t, \ell_{j+1}), t}+\frac{|NL(T_{(A(p_t, \ell_{j+1}))})|}{k})^{-1}-(u_{A(p_t, \ell_j), t}+\frac{|NL(T_{(A(p_t, \ell_j))})|}{k})^{-1}]$

\hspace{1.3cm}$=\frac{u_r+\frac{|S|}{k}}{\phi({\ell_j+1})}\cdot [(u_{A(p_t, \ell_j+1), t}+\frac{|NL(T_{(A(p_t, \ell_j+1))})|}{k})^{-1}-(u_{A(p_t, \ell_j), t}+\frac{|NL(T_{(A(p_t, \ell_j))})|}{k})^{-1}]$

\end{proof}
The proof for Theorem 4.7 is as follows:

\begin{proof}

Let $P$ denote the value of the objective function of the primal
solution and $D$ denote the value of the objective function of the
dual solution. Initially, let $P=0$ and $D=0$. In the following, we
prove three claims:

(1) The primal solution produced by the algorithm is feasible.

(2)  The dual  solution produced by the algorithm is feasible.

(3) $ P\leq 2\ln(1+k)D$.

By three claims and weak duality of linear programs, the theorem
follows immediately.

The proof of claim (1) and (2) are similar to that of claim (1) and (2) in section 3.7.

Third, we prove claim (3) as follows.  If  the algorithm increases
the variables $a_{S,t}$ at some time $t$, then: $\frac{\partial
D}{\partial a_{S,t}}=|S|-k$. Let's compute the primal cost. At depth $j\in \{\ell_1+1, \ell_2+1, \ldots, \ell_h+1\}$, we compute the movement cost of our algorithm by the change of $\nabla b'_j$ as follows.

%


$\sum\limits_{w\in NFC(A(p_t, j-1))\backslash
\{A(p_t, j)\}}2D(w)\cdot \frac{du_{w,t}}{db_j'}\cdot \frac{db_j'}{a_S}$

$\hspace{3.0cm}=\nabla b'_j\sum\limits_{w\in C(A(p_t, j-1))\backslash
\{A(p_t, j)\}}2D(w)\cdot\frac{\ln(1+k)}{2D(w)}(u_{w,t}+\frac{|NL(T_w)|}{k})$

$\hspace{3.0cm}=\frac{u_{r, t}+\frac{|S|}{k}}{\phi(j)}\cdot\ln(1+k)\cdot\frac{\sum\limits_{w\in C(A(p_t, j-1))\backslash
\{A(p_t, j)\}}(u_{w,t}+\frac{|NL(T_w)|}{k})}{u_{A(p_t, j-1),t}+\frac{|NL(T_{A(p_t, j-1)})|}{k}}$

$\hspace{3.0cm}=\frac{\sum\limits_{p\in S}u_{p,t}+\frac{|S|}{k}}{\phi(j)}\cdot\ln(1+k)\cdot\frac{\sum\limits_{w\in C(A(p_t, j-1))\backslash
\{A(p_t, j)\}}(u_{w,t}+\frac{|NL(T_w)|}{k})}{u_{A(p_t, j-1),t}+\frac{|NL(T_{A(p_t, j-1)})|}{k}}$

$\hspace{3.0cm}<\frac{(\sum\limits_{p\in S\setminus\{p_t\}}u_{p,t}+\frac{|S|-1}{k}+2)}{\phi(j)}\cdot\ln(1+k)\cdot\frac{\sum\limits_{w\in C(A(p_t, j-1))\backslash
\{A(p_t, j)\}}(u_{w,t}+\frac{|NL(T_w)|}{k})}{u_{A(p_t, j-1),t}+\frac{|NL(T_{A(p_t, j-1)})|}{k}}$

$\hspace{3.0cm}=\frac{(\sum\limits_{p\in S\setminus\{p_t\}}u_{p,t}+\frac{|S|-1}{k}+2(|S|-k))}{\phi(j)}\cdot\ln(1+k)\cdot\frac{\sum\limits_{w\in C(A(p_t, j-1))\backslash
\{A(p_t, j)\}}(u_{w,t}+\frac{|NL(T_w)|}{k})}{u_{A(p_t, j-1),t}+\frac{|NL(T_{A(p_t, j-1)})|}{k}}$

$\hspace{3.0cm}\leq (3(|S|-k)+\frac{(|S|-1)}{k})\cdot\ln(1+k)\cdot \frac{\sum\limits_{w\in C(A(p_t, j-1))\backslash
\{A(p_t, j)\}}(u_{w,t}+\frac{|T(w)|}{k})}{u_{A(p_t, j-1),t}+\frac{|NL(T_{A(p_t, j-1)})|}{k}}$

$\hspace{3.0cm}\leq 4\cdot\ln(1+k)\cdot (|S|-k)\cdot \frac{\sum\limits_{w\in C(A(p_t, j-1))\backslash
\{A(p_t, j)\}}(u_{w,t}+\frac{|NL(T_w)|}{k})}{u_{A(p_t, j-1),t}+\frac{|NL(T_{A(p_t, j-1)})|}{k}}$

$\hspace{3.0cm}\leq 4\cdot\ln(1+k)\cdot(|S|-k)$, since $\sum\limits_{w\in C(A(p_t, j-1))\backslash
\{A(p_t, j)\}}(u_{w,t}+\frac{|NL(T_w)|}{k})\leq u_{A(p_t, j-1),t}+\frac{|NL(T_{A(p_t, j-1)})|}{k}$.

%
%
%
%
%
%
%
%
%
%
%
%
%
%

Where the first inequality holds since $\sum\limits_{p\in S\setminus\{p_t\}}u_{p,t}<
|S|-k$, the reason is that the constraint at time $t$ is not
satisfied otherwise the algorithm stop increasing the variable
$u_{p,t }$ (Since $\sum\limits_{p\in S}u_{p,t}=|S|-k$, $\sum\limits_{p\in S\setminus\{p_t\}}u_{p,t}<
|S|-k\Leftrightarrow u_{p_t, t}=0$, i.e the algorithm stop increasing the variables). In addition, when $|S|\geq k+1,\frac{|S|-1}{k}= |S|-k$.

Thus, the total cost of all $j$ depth is at most $4\ell\cdot \ln (1+k)\cdot(|S|-k)$.
Hence, we get $ P\leq 4\dot \ell\cdot \ln(1+k)\cdot D$.


So, the competitive ratio of this algorithm is  $4\ell\ln(1+k)$.

%
%
%
%
%

\end{proof}

\end {document}